\documentclass[3p,times,twocolumn]{elsarticle}
\usepackage{amssymb}
\usepackage{amsfonts}
\usepackage{amsmath}
\usepackage{graphicx}
\usepackage{dcolumn}
\usepackage{soul}
\usepackage{bm}
\usepackage{revsymb}
\usepackage{appendix}
\usepackage{xcolor}%
\usepackage{ulem}
\begin{document}
\begin{frontmatter}
\title{Topologically Protected Spatially Localized Modes: An Easy Experimental Realization of the Su--Schrieffer--Heeger Model}
\author{L.Q. English}
\address{Department of Physics and Astronomy, Dickinson College, Carlisle, PA 17013, USA}
\author{A. Halchenko}
\address{Department of Physics and Astronomy, Dickinson College, Carlisle, PA 17013, USA}
\author{F. Palmero}
\address{Grupo de F\'{\i}sica No Lineal, Departamento de F\'{\i}sica Aplicada I, Escuela T\'{e}cnica Superior de Ingenier\'{\i}a Inform\'{a}tica, Universidad de Sevilla, Avda Reina Mercedes s/n, E-41012 Sevilla, Spain}
\ead{palmero@us.es}

\begin{abstract}
In this paper, we review the basic concepts of topologically protected edge modes using the Su–Schrieffer–Heeger (SSH) model, originally introduced to describe electrical conductivity in doped polyacetylene polymer chains. We then propose an electrical circuit that emulates this model, provide its mathematical description, and present its experimental realization. The experimental setup is described in detail, with explanations designed to be broadly accessible without much prior familiarity with lattice theory, thus offering an introduction to this active area of research. Both theoretical predictions and experimental results confirm the presence of these modes, showing very good overall agreement. Using this concrete experimental system as a motivating example, we highlight the key aspects of topological protection.
\end{abstract}

\begin{keyword}
	Topological modes \sep ssh model \sep electrical circuits
\end{keyword}

\end{frontmatter}

\section{Introduction}
While topological phases of matter is not a new subject, going back to the discovery of topological insulators, they have in recent years seen increasing interest due to their central promise in emerging technologies, such as in quantum computation, where the feature of topological protection holds the promise of robust Q-bits. More generally, topologically protected modes have become a paradigm in the study of localized wave propagation in condensed-matter and photonic systems. Topological protection can stabilize and localize signal propagation \cite{Sou17, Woo18, Sha22}, and has consequently spawned its own active area of research \cite{Haf11, Asb12, Rec13, Rec13b, Lu14, Asb14, Ver15, Oza19, Has19, PA24}.  

Perhaps the simplest model featuring topological modes is the Su–Schrieffer–Heeger (SSH) model, originally proposed to describe the electrical conductivity of a doped polyacetylene polymer chain \cite{SSH79}. This polymer is essentially a one-dimensional hydro-carbon chain characterized by an alternation of single and double bonds along the chain. 

In this paper, we begin with a basic review of topologically protected modes, their origin and significance, using this simple one-dimensional model, and we propose and realize an electrical lattice built from inexpensive and easily accessible components that emulates the SSH system. The experimental realization of topological edge states in such lattices forms the thrust of the paper, and we investigate and describe in detail their existence, generation, and properties; we also investigate bound states at domain walls that emerge in this system. Our system allows for a full spatial mapping of such modes in this one-dimensional lattice. Two-dimensional topological circuits have also been studied \cite{imhof, ningyuan, hofmann} in which access to spatial information, however, tended to be more limited or indirect.    

Furthermore, by introducing long-range coupling, we show how we can reach novel topological phases that go beyond the standard SSH model within certain well-defined parameter regimes. These phases feature new topological modes, and we show how to generate these in our system. In all cases,
we find good agreement between theoretical predictions and experimental data, which both confirm the presence of these nontrivial modes.

From a larger perspective, there are several advantages of having a electrical-lattice emulation of the famous SSH model. One is that it can help make the model and its mathematical features more concrete and accessible, thus serving as an ideal introduction to it. The other is that the context of electrical lattices provides versatility and adaptability in design and allows for the flexible implementation of specific lattice geometries and parameter regimes of interest.

\section{The SSH Model: From Theoretical Concepts to Experimental Realization}
\begin{figure}[h]
\centering
\includegraphics[width=\columnwidth]{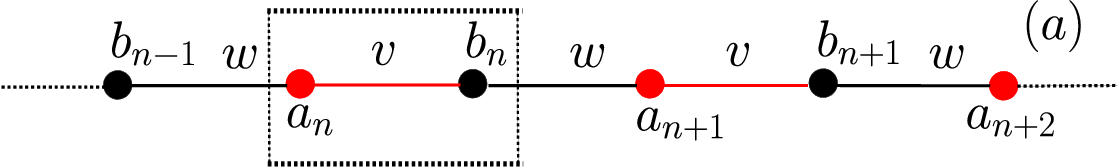}
\includegraphics[width=\columnwidth]{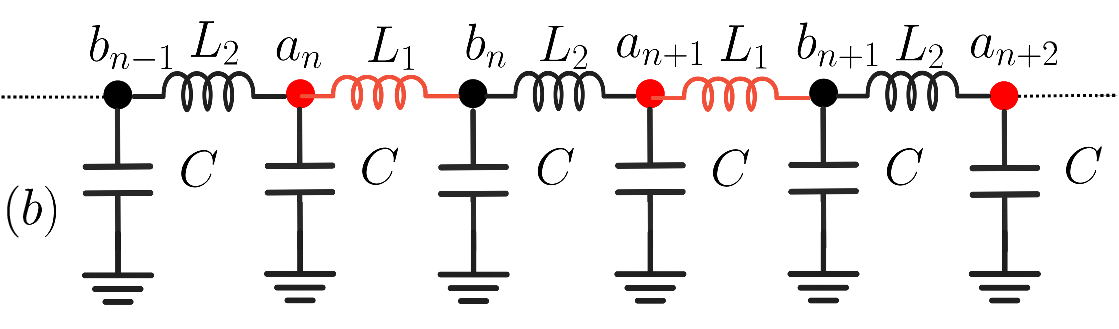}
\caption{(a) Sketch of the SSH model, where the unit cell $(a_n,b_n)$ is in dotted box. Coupling constant is $v$ within the unit cell and $w$ between cells. (b) Sketch of the equivalent electric lattice.}
\label{SSH}%
\end{figure}

Although the SSH model was originally formulated as a quantum model to describe the hopping of spinless fermions (tight-binding model), it can also be regarded as a one-dimensional chain of oscillators formed by a periodic repetition of two-site unit cells, with two different coupling strengths, \(v\) and \(w\), within and between the cells, respectively, as shown in Fig.~(\ref{SSH} a).


In order to build a real system that mimics the SSH model, and following \cite{Chi18,Tej21,Anz25}, we construct an electrical lattice: a one-dimensional transmission line in which a set of $N$ capacitors are coupled through inductors, as shown in Fig.~\ref{SSH}(b). The coupling alternates between two different inductances, forming two-site unit cells, or two sublattices, $a_n$ and $b_n$.

Applying Kirchhoff's law, the equations describing the circuit read as 
\begin{align} 
    \frac{d^2 a_n}{d t^2} & = v (b_n-a_n)+w(b_{n-1}-a_n),  \label{in_latt_1} \\
    \frac{d^2 b_n}{d t^2} & = v (a_n-b_n)+ w(a_{n+1}-b_n),
   \label{in_latt_2}
\end{align}
where $n=1 \dots N$, $v = 1/L_1 C$, $w = 1/L_2 C$, and $(a_n$, $b_n)$  are the voltages at the lattice node $n$. The resistance of the elements was measured and found to be negligible, allowing us to disregard dissipative effects. As a first step, we consider a ring lattice of $2N$ sites, or equivalently, $N$ unit cells, and  periodic boundary conditions, so that the last site is connected back to the first. Other configurations have also been used, as discussed below. 

\begin{figure}[h]
\centering
\includegraphics[width=\columnwidth]{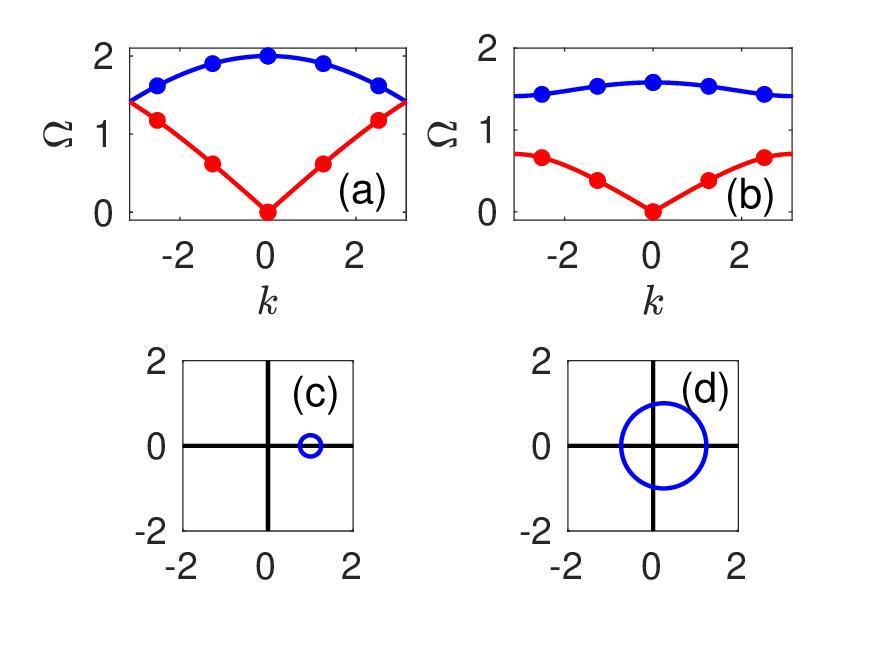}
\caption{(a) and (b) Band structure. (a) \(v = w = 1\), (b) \(v \neq w\) (\(v = 0.25\) and \(w = 1\)). The continuous blue line corresponds to the optical band in an infinite lattice, and the red line to the acoustic band (in an infinite lattice). Solid points correspond to the frequencies of the dispersion bands for a finite lattice of \(N = 5\) unit cells with periodic boundary conditions. The same bands are obtained when \(v = 1\) and \(w = 0.25\), i.e., when the coupling strengths are swapped. Panels (c) and (d) show the circle in the complex plane described by \(h(k)\) as \(k\) varies between \(-\pi\) and \(\pi\). (c) \(v = 1\) and \(w = 0.25\), (d) \(v = 0.25\) and \(w = 1\).}
\label{band_new}%
\end{figure}

We look for plane-wave solutions of the form:
\begin{align}
    a_n(t)=a e^{i(k n-\Omega t)}  \label{an} \\
    b_n(t)=b e^{i(k n-\Omega t)}. \label{bn} 
\end{align}
These solutions are periodic in \(k\) with period \(2\pi\), so we can restrict our study to the so-called first Brillouin zone, which consists of the set of \(k\) points close to the origin in the interval \((-\pi,\pi)\). In the case of a ring structure, the periodicity of the lattice implies that \(k\) is quantized, namely \(k = 2\pi n / N\), with \(n = 0, \dots, N-1\). Substituting Eqs.~(\ref{an},\ref{bn}) into Eqs.~(\ref{in_latt_1}, \ref{in_latt_2}), the problem is transformed into an eigenvalue problem, $H(k)\, \mathbf{v} = \Omega^2 \mathbf{v}$, where \(\mathbf{v} = (a,b)^{T}\) and \(H(k)\) is a Hermitian \(2 \times 2\) matrix.

\begin{equation}
H(k)=\begin{pmatrix}
v+w  & h(k) \\
h^*(k) & v+w
\end{pmatrix},
\label{H_k}
\end{equation}
with $h(k)=-(v+w e^{-i k})$. The eigenvalues are then evaluated as,
\begin{equation}
    \Omega^2(k)=v+w \pm \sqrt{v^2+w^2+2 v w \cos k},
    \label{disp}
\end{equation}
This gives a relation between the frequency and the wavenumber \(k\), known as the dispersion relation, where each frequency corresponds to a value of the wavenumber. In an infinite lattice, \(k\) can take any value between \(-\pi\) and \(\pi\), so the corresponding frequencies form what is called a band. The minus sign corresponds to the acoustic band (the minimum frequency is zero), while the plus sign corresponds to the optical band. The band structure is shown in Fig.~\ref{band_new} (a) and (b). In a finite lattice we will get only a set of $N$ points corresponding to each band.

When $v \neq w$, a band gap appears, which closes when the two coupling strengths become equal. Owing to the symmetry of the dispersion relation, the dispersion relation is unchanged when $v$ and $w$ are swapped, which might suggest that both situations are identical. However, this is not the case. It turns out that the two situations are topologically distinct.

In order to illustrate this difference we will reproduce a similar argument as shown in \cite{PA24}. The problem of finding the planar waves of our system reduces to diagonalizing the Hermitian matrix of Eq.(\ref{H_k}). The eigenvalues therefore yield $\Omega$ and recover the dispersion relation in Eq.~(\ref{disp}). After some more work, the normalized eigenvectors can also be found; they are given by,
\begin{align}
    {\bf v}_{\pm}(k)&= \frac{1}{\sqrt{2}} \begin{pmatrix}
           \pm \hat{h}(k) \\
           1 \\
         \end{pmatrix},
         \label{v}
\end{align}
where 
\begin{align}
\hat{h}(k) = \frac{u + v e^{-ik}}{\sqrt{u^2 + v^2 + 2 u v \cos k}}.
\end{align}
The argument of the complex number $\hat{h}(k)$ is the same as that of $h(k)$. As the wavenumber $k$ varies from $-\pi$ to $\pi$, the number $h(k)$ traces a circle in the complex plane with center at $v$ and radius $w$. This circle encloses the origin when $v < w$ (topological phase), but not when $v > u$ (trivial phase), as shown in Fig.~\ref{band_new} (c) and (d). The number of times this complex number encircles the origin can be defined as the winding number, a topological invariant, which is $-1$ in the first case and $0$ in the second. 

The two situations are fundamentally different. Transforming continuously the system 
from one to the other requires crossing the origin point \((v = w)\), which 
corresponds to a topological phase transition, a type of phase change in which the 
fundamental topological properties of a system change abruptly, giving rise to distinct 
phases that are indistinguishable locally but different globally, as in the case of a change in the number of holes of an object. From a physical perspective, this means transitioning from a typical band structure characteristic of an insulator to another insulating phase, but passing through a critical point where the band gap closes. At this point, the band structure temporarily corresponds to that of a conductor.  For a deeper study of this subject and a rigurous discussion about these topics see  \cite{Jan16}. In a infinite or ring lattice, one could go from one case to the other simply by shifting the unit cell over by half a lattice spacing. However, this does not work in the presence of lattice edges. 

\subsection{Topological edge modes}
Previously, we focused on plane-wave modes in an infinite lattice or in a finite lattice with periodic boundary conditions. In a large but finite lattice, these modes correspond to {\it bulk modes}, namely, states that live in the interior (bulk) of the lattice rather than at its boundaries or edges. In general, there exists a correspondence between bulk and edge states in a lattice with free boundaries (the {\it bulk--edge correspondence}): whenever a system possesses nonzero topological invariants, it supports the existence of edge states that are, in general, ``robust'' against perturbations. Although a fully general proof of this correspondence does not exist, rigorous proofs have been established for systems of this type~\cite{Jan16}.

\begin{figure}[h]
\centering
\includegraphics[width=\columnwidth]{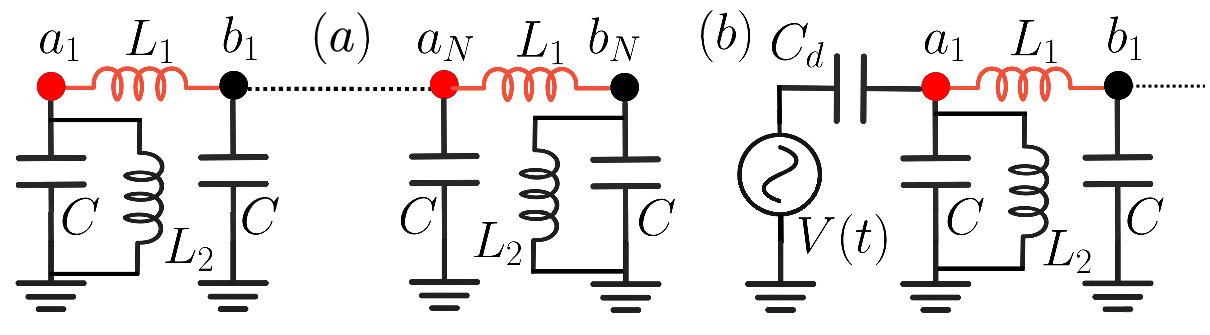}
\caption{(a) Sketch of the ends of a finite lattice. Note that, in order to keep structure of the equations, we have to include extra inductors $L_2$ connected to the ground in both ends. (b) The lattice is driven sinusoidally at one end via a capacitor $C_d$ in order to induce the linear modes.}
\label{lattice_ends}%
\end{figure}

In our system, this characteristics is evident when we analyze a finite lattice of $N$ unit cells ($2N$ nodes), where the two ends are shown in Fig. (\ref{lattice_ends}). To preserve the symmetry of the system, additional inductors $L_2$ are included, each connected to ground at the two ends of the chain.

 Equations corresponding to the two ends can be derived from Eqs.(\ref{in_latt_1}, \ref{in_latt_2}) by realizing that there is no site to the left of $n=1$ and no site to the right of $n=N$. Therefore, we see that the amplitudes of the linear modes corresponding to the finite lattice will be the eigenvectors of the matrix: 
\begin{equation}
H = \begin{pmatrix}
0 & v & 0 & \cdots & 0 \\
v & 0 & w & \ddots & \vdots \\
0 & w& 0 &v \ddots & 0 \\
\vdots & \ddots & \ddots & \ddots &v \\
0 & \cdots & 0 & v & 0
\end{pmatrix},
\label{H}
\end{equation}
and the eigenvalues are the frequency combination $v + w - \Omega^2$.

Topological edge modes arise when $v<w$. To gain insight into this situation, we analyze two extreme cases~\cite{Jan16}. If $w = 0$, the lattice reduces to a set of $N$ independent dimers coupled through $v$, and no edge modes appear. In contrast, if $v = 0$, there exist two uncoupled modes at the ends of the lattice, which give rise to the localized edge modes.

\begin{figure}[h]
\centering
\includegraphics[width=\columnwidth]{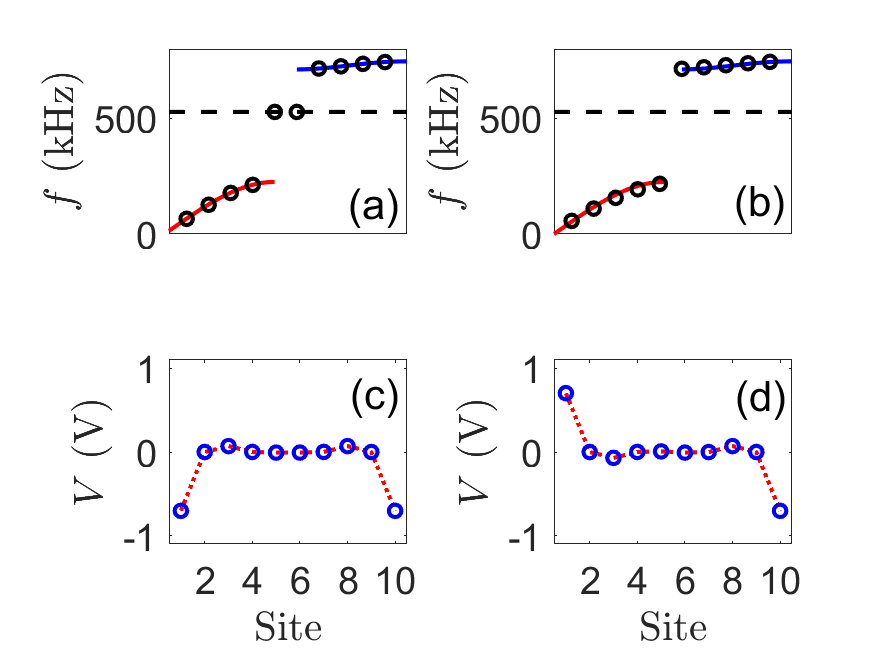}
\caption{Numerical Frequencies and localized edge modes calculated numerically diagonalizing the matrix $H$. (a) Frequencies (black circles), where a piece of the optical (blue  line) and acoustic (red  line) corresponding to the infinite lattice has been superimposed. In dashed black line we show the frequency corresponding to $\sqrt{v+w}$ (in kHz). (b) Same as (a), but with the frequencies switched ($v>w$). In this case, no edge modes appear in the lattice. (c) and (d) show the localized edge modes profile (blue circles) and the approximate solutions determined by (\ref{m1}) and (\ref{m2}) (red dotted line).  Electric lattice where $L_1=1$ mH, $L_2=0.1$ mH, $C=1$ nF and $N=5$.}
\label{frec_edge}%
\end{figure}

There are two edge modes with frequencies in the band gap, as shown in Fig. (\ref{frec_edge}). We can obtain an expression for these edge modes for a semi-infinity chain, which remains  valid for a finite chain when \((v / w)^{(N-1)} \approx 0\). Modes have a frequency $\Omega = \sqrt{v+w}$, and are given by any linear combination of the two (normalized) independent vectors $\vec{s}_1$ and $\vec{s}_2$:
\begin{align}
    \vec{s}_1&= \begin{pmatrix}
           x_n \\
           y_n \\
         \end{pmatrix} =C \begin{pmatrix}
           1 & \cdots & -(v/w)^{n-1} & \cdots & -(v/w)^{N-1} \\
           0 & \cdots & 0 & \cdots & 0\\
         \end{pmatrix}, \label{m1}\\
\vec{s}_2&= \begin{pmatrix}
           x_n \\ 
           y_n \\
         \end{pmatrix} =C \begin{pmatrix}
          0 & \cdots & 0 & \cdots & 0\\
         (-v/w)^{N-1} & \cdots & (-v/w)^{N-n} & \cdots & 1
         \end{pmatrix},
         \label{m2}
\end{align}
where $C$ is a normalization constant, and $x_n$ and $y_n$ denote the spatial components of the solutions $a_n(t)$ and $b_n(t)$, respectively. As shown in Fig.~\ref{frec_edge}, these expressions provide good approximations even when the lattice is not very large or when the frequency ratio is not particularly small.

One interesting feature of Fig.~\ref{frec_edge} is that the number of edge modes is either zero or two; in other words, they are created in pairs as we move from the trivial phase to the topological phase, and that within that pair, the only difference is that in sign on one sublattice  is reversed but the sign of the other sublattice is preserved when comparing the two modes. This can be understood as a consequence of a type of symmetry inherent in the system, called {\it chiral symmetry}.

In fact, the matrix $H$ in Eq.(\ref{H}), anti-commutes with the matrix,
\begin{equation}
\label{gamma}
\Gamma = \begin{pmatrix}
1 & 0 & 0 & \cdots & 0 \\
0 & -1 & 0 & \ddots & \vdots \\
0 & 0& 1 &0 \ddots & 0 \\
\vdots & \ddots & \ddots & \ddots &0 \\
0 & \cdots & 0 & 0 & -1
\end{pmatrix},
\end{equation}
which inverts values on sublattice B, while perserving values on sublattice A. In other words, it is easy to show that,

\begin{equation}
    H \Gamma = - \Gamma H.
    \label{anticom}
\end{equation}
Consequently, if  \[\vec{t}_1= \begin{pmatrix}
           x_n \\
           y_n \\
         \end{pmatrix}\] 
is an eigenvector of $H$ with an eigenvalue of $\lambda_1$, then it there must also exist a second eigenvector,
\[\vec{t}_2 = \begin{pmatrix}
           x_n \\
           -y_n \\
         \end{pmatrix}\]
with an eigenvalue of $\lambda_2=-\lambda_1$.

What is interesting is that an edge mode can be shown to exist at $\lambda=0$, so there must then be a degenerate partner at that same eigenvalue. This is exactly what is observed in Fig.~\ref{frec_edge}. Notice that we can now form a linear combination of these two eigenvectors (a sum) such that all entries on sublattice $B$ are zero, and another combination (a difference) such that all entries on sublattice $A$ are zero, and these two cases will result in edge modes that live either entirely on the left edge or entirely the right edge of the lattice. This scenario is given in Eqs.(\ref{m1}-\ref{m2}).

Furthermore, these edge states are robust against changes in lattice parameters. This can also be understood via symmetry arguments. The chiral symmetry, encapsulated in Eq.(\ref{anticom}), keeps the two eigenvalues degenerate at $\lambda=0$. This is because if one parameter change caused $\lambda_1$ to become positive, the other one, $\lambda_2$, would have to become negative. This opposite frequency shift, however, should not happen ordinarily because the system also has a left-right symmetry under reflection about the midpoint. Therefore, the left-edge localized mode and the right-edge localized mode would be expected to have identical frequencies. Indeed, we can numerically verify that in long lattices, the degeneracy of the pair of edge modes is difficult to lift - hence their robustness.

To study experimentally these edge modes we have built a lattice of $10$ capacitors of capacitance $C=1.0$ nF  connected to the ground ($N=5$) and coupled by two sets of inductors, $L_1$ and $L_2$.  Both ends of the lattice are connected to the ground through inductors $L_2$. Ignoring the (small) dissipative effects due to the resistance of elements, mathematically the system can be described by Eqs. (\ref{in_latt_1}, \ref{in_latt_2}). In our experiments, we have used two sets of inductors for $L_1$ ($L_1=1$ mH and $L_1=0.470$ mH), and inductors $L_2=0.1$ mH.

To experimentally detect linear modes we supply energy locally at one end via a driving capacitor $C_d=40$ pF ($Cd << C$)  and a sinusoidal voltage input from a signal generator (Agilent 33220A function/sweep generator), as shown in Fig.\ref{lattice_ends}(b). Spectra are recorded by sweeping the drive frequency with a constant driven amplitude $V_d \approx 1$ V, and measuring the voltage of a particular site with an oscilloscope. Once the driver frequency coincides with a resonance frequency corresponding to a linear mode, the increase in response amplitude is recoded by the oscilloscope, and spectra are thus obtained. These spectra will depend on which site is driven and observed. To ascertain the spatial profile of any mode in the spectrum, we tune to function generator to that frequency while simultaneously monitoring the voltages at all lattice sites using a 16-channel data acquisition system (NI PXI-1033 with NI 6133 cards). (In principle, this task could also be done sequentially with an oscilloscope.)

\begin{figure}[h]
\centering
\includegraphics[width=\columnwidth]{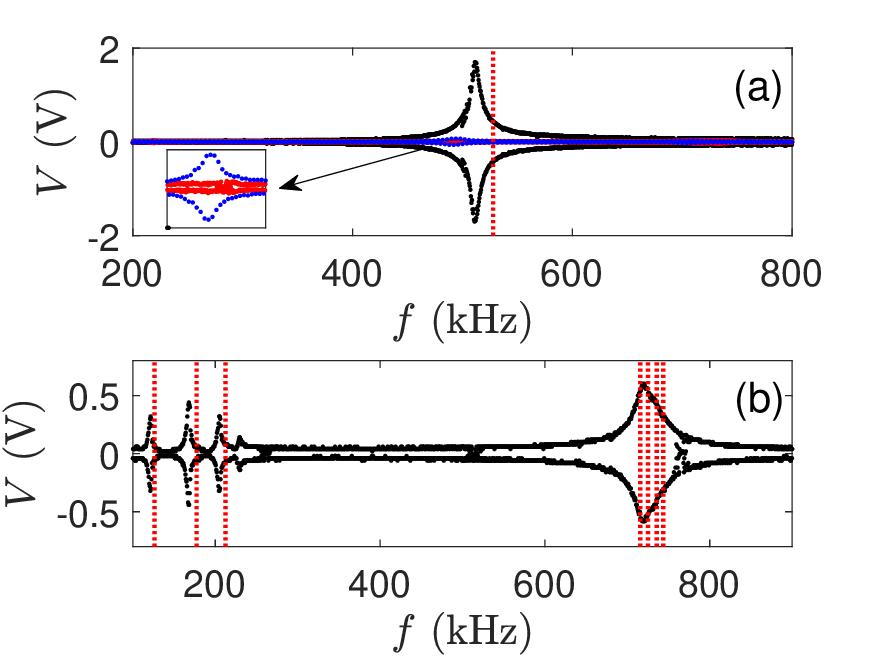}
\caption{Experimental voltage at the driven end and its neighboring sites as a function of the driving frequency when energy is supplied locally at the end by a sinusoidal voltage source with amplitude $V_d = 1 \,\text{V}$. Circuit parameters: $L_1 = 1 \,\text{mH}$, $L_2 = 0.1 \,\text{mH}$, and $C = 1 \,\text{nF}$.
(a) Black points represent the voltage at the end node, red points the voltage at its first neighbor, and blue points the voltage at its second neighbor (see inset). 
(b) Voltage response at the second node. Dashed red lines indicate the theoretical resonance frequencies corresponding to edge modes in (a) and to acoustic and optical modes in (b).}
\label{osc_1}%
\end{figure}

By measuring the voltage at the driven end and its neighboring site, we obtain the spectrum shown in Fig.~\ref{osc_1}(a). A strong response is observed at $f = 511$ kHz in the end node, a weak response at the second neighbor while the rest of the lattice shows almost no response. This result shows a good agreement with theoretical predictions where the frequency corresponding to edge nodes is $f=554$ kHz. The frequency shift can be explained by 
 the effect of the small capacitor used to connect the voltage source to the lattice, to parasitic inductance introduced by the ribbon cables, and parasitic capacitance due to the acquisition channels. When monitoring the voltage of the first neighbor (see Fig.~\ref{osc_1}(b)), resonances corresponding to acoustic and optical modes are observed. Comparable results were obtained for the other set of inductors.

Interestingly, when we modify the end of the lattice by adding an additional unit cell and starting with the small inductor, thus making the first coupling stronger than the second, $v>w$, we do not observe any resonances corresponding to edge modes. The peak in Fig.(~\ref{osc_1}a) vanishes entirely. In other words, the edge mode only exists experimentally when $v<w$, as expected theoretically.

\begin{figure}[h]
\centering
\includegraphics[width=\columnwidth]{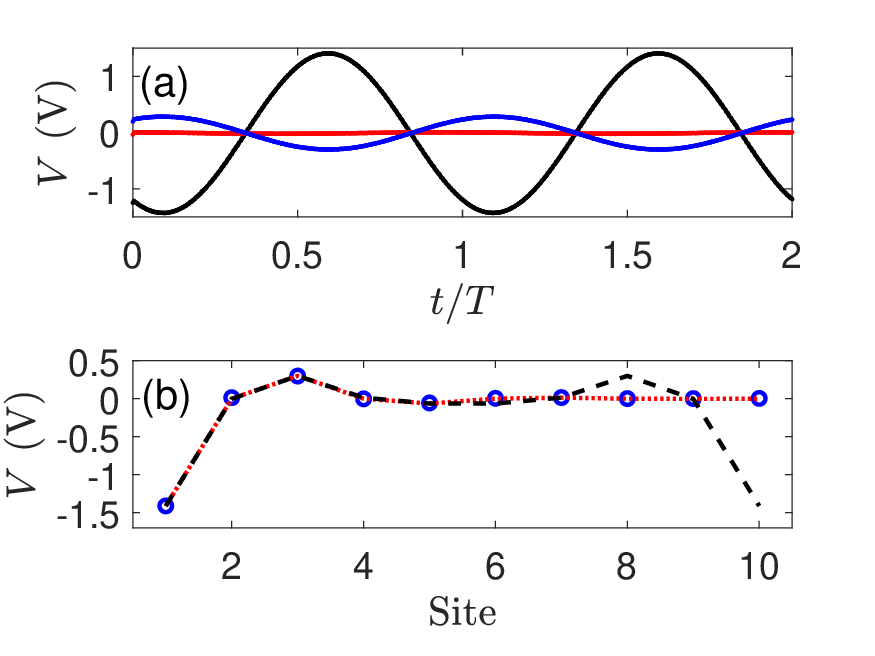}
\caption{(a) Experimental voltage as function of time at the driven end (black line) and its first (red line)  and second neighbor (blue line) corresponding to the edge mode for a frequency of $f=554$ kHzs. (b) Edge mode profile. Circles correspond to experimental measurements, dashed black line to normalized numerical calculations and dotted red line to normalized approximate solutions determined by (\ref{m1}) and (\ref{m2}). Circuit parameters: $L_1 = 0.470 \,\text{mH}$, $L_2 = 0.1 \,\text{mH}$, and $C = 1 \,\text{nF}$.}
\label{edge_1}%
\end{figure}

In Fig.~(\ref{edge_1}), 
we show the experimental and theoretical results corresponding to the edge mode. Since we are driving the lattice at only one end, we can observe the localized mode at that end only. In other words, we are exciting a superposition of the degenerate even and odd edge states predicted by theory.

Figure \ref{edge_1}(a) shows the voltage response in time at the first three lattice sites (black, blue, red), and (b) depicts the voltage profile spatially at a given snapshot in time. Here, the blue circles are experimental data-points, and the lines (black, red) show the numerical and analytical predictions, respectively. We observe good agreement between all three. Repeating these measurements with a smaller \( L_1/L_2 \) ratio---thereby producing more weakly localized edge modes---also yields excellent agreement with both the numerical and analytical predictions.

\subsection{Long range coupling: How to manipulate the number of topological edge modes}
Given the preceding discussion, it is clear that chiral symmetry plays a central role in the existence of topological edge modes. It is indeed possible to introduce longer-range couplings while preserving this symmetry, but care must be taken regarding their structure \cite{Li24}. In particular, adding couplings between sites belonging to the same sublattice does \emph{not} work, as can be readily verified numerically.

\begin{figure}[h]
	\centering
	\includegraphics[width=\columnwidth]{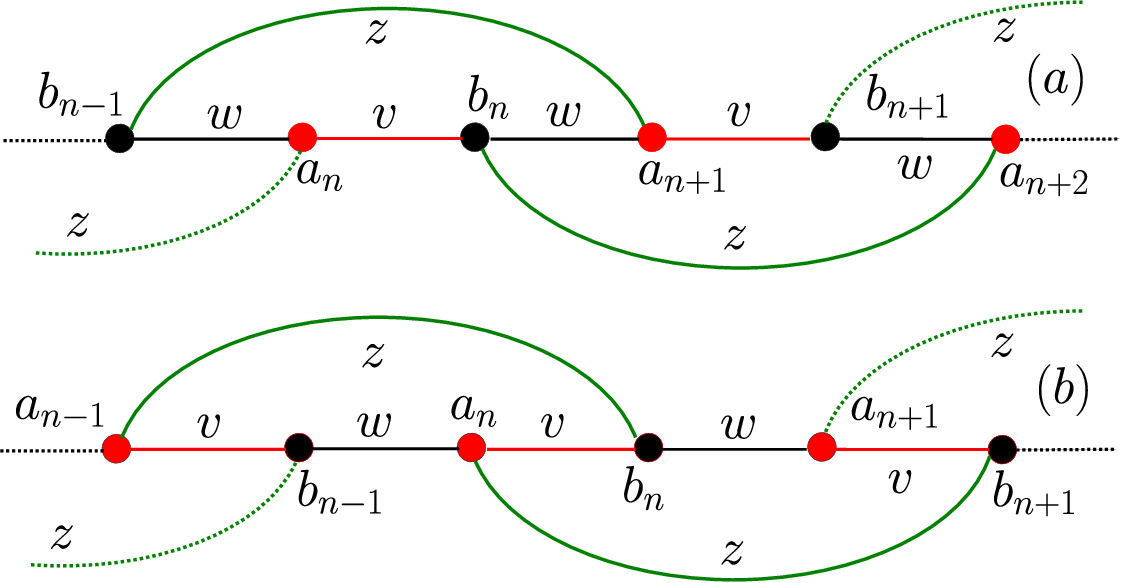}
	\caption{Sketch of the electrical lattice with long-range coupling. 
		(a) Configuration in which an extra inductor (in green) between the B-site of the \(n^{\text{th}}\) unit cell and the A-site of the \((n+2)^{\text{th}}\) unit cell has been included. 
		(b) The extra inductor (in green) is now placed between the A-site of the \(n^{\text{th}}\) unit cell and the B-site of the \((n+1)^{\text{th}}\) unit cell.}
	\label{SSH_4}
\end{figure}

One way to incorporate symmetry-preserving longer-range interactions is illustrated in Fig.~\ref{SSH_4}. In panel (a), we add inductors between the B-site of the \(n^{\text{th}}\) unit cell and the A-site of the \((n+2)^{\text{th}}\) unit cell. A second approach, shown in panel (b), is to connect the A-site of the \(n^{\text{th}}\) unit cell to the B-site of the \((n+1)^{\text{th}}\) unit cell. In both cases, the additional coupling has impedance \(z = 1/(L_3 C)\), where \(L_3\) is the inductance of this third inductor.

For a lattice with \(N = 5\), these schemes introduce three extra inductors. Implementing them also requires adding compensating inductors to ground, ensuring that every lattice node remains connected to exactly three inductors.

Using matrix-multiplication software, it is not difficult to show that the new matrix (analogous to Eq.(\ref{H}) but with additional $z$-entries) also anti-commutes with the symmetry matrix $\Gamma$, Eq.~(\ref{gamma}), and so it preserves chirality. 

Moreover, the introduction of these next-neighbor couplings can modify an essential topological property of the lattice—the winding number—which can be computed analytically \cite{Jan16, Li24}. In the first case, where a coupling is added between the B-site of the \(n^{\text{th}}\) unit cell and the A-site of the \((n+2)^{\text{th}}\) unit cell, it can be shown that when \(z > v > w\), the winding number becomes \(\nu = -2\). This implies the existence of two pairs of topological edge states.

\begin{figure}[h]
	\centering
	\includegraphics[width=\columnwidth]{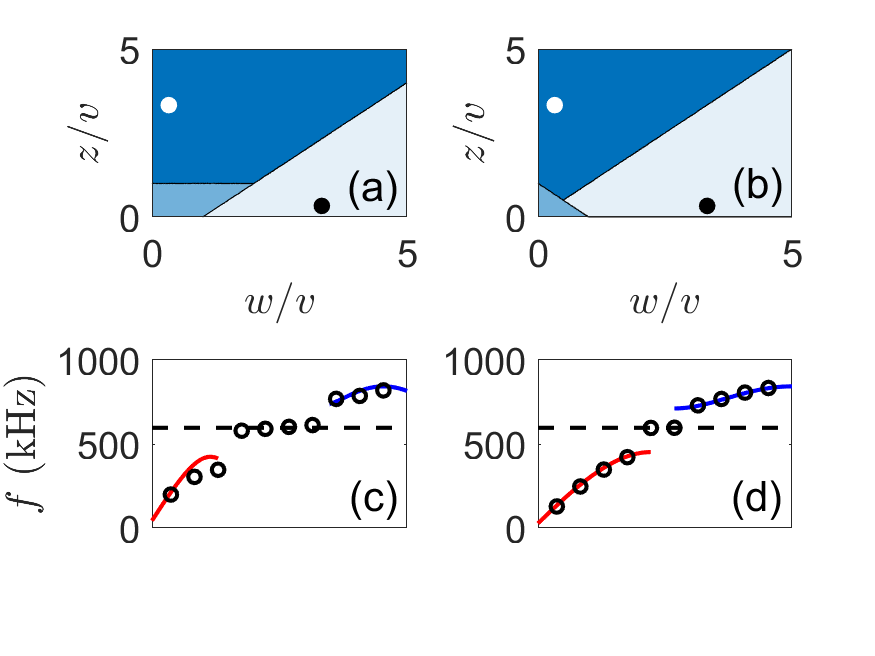}
		\caption{Phase diagrams representing the winding number $\nu$ as a function of the coupling parameters. Panel (a) corresponds to a long-range coupling between the B-site of the $n^{\text{th}}$ unit cell and the A-site of the $(n+2)^{\text{th}}$ unit cell, while panel (b) corresponds to coupling between the A-site of the $n^{\text{th}}$ unit cell and the B-site of the $(n+1)^{\text{th}}$ unit cell. In (a), the dark blue region corresponds to $\nu = -2$, the light blue region to $\nu = -1$, and the medium blue region to $\nu = 0$. In (b), the dark blue region corresponds to $\nu = 1$, the light blue region to $\nu = -1$, and the medium blue region to $\nu = 0$. Panels (c) and (d) show typical band structures corresponding to cases where the winding number is $2$ and $1$, respectively. Continuous lines represent portions of the dispersion relation for an infinite lattice, while circles denote numerical calculations for a finite lattice with $N = 5$ unit cells (10 sites). The dashed black line indicates the frequencies associated with edge modes. In (a) and (b), black (white) points mark the parameter values that have been studied experimentally.}
	\label{phase_lr}%
\end{figure}

\begin{figure}[h]
	\centering
	\includegraphics[width=\columnwidth]{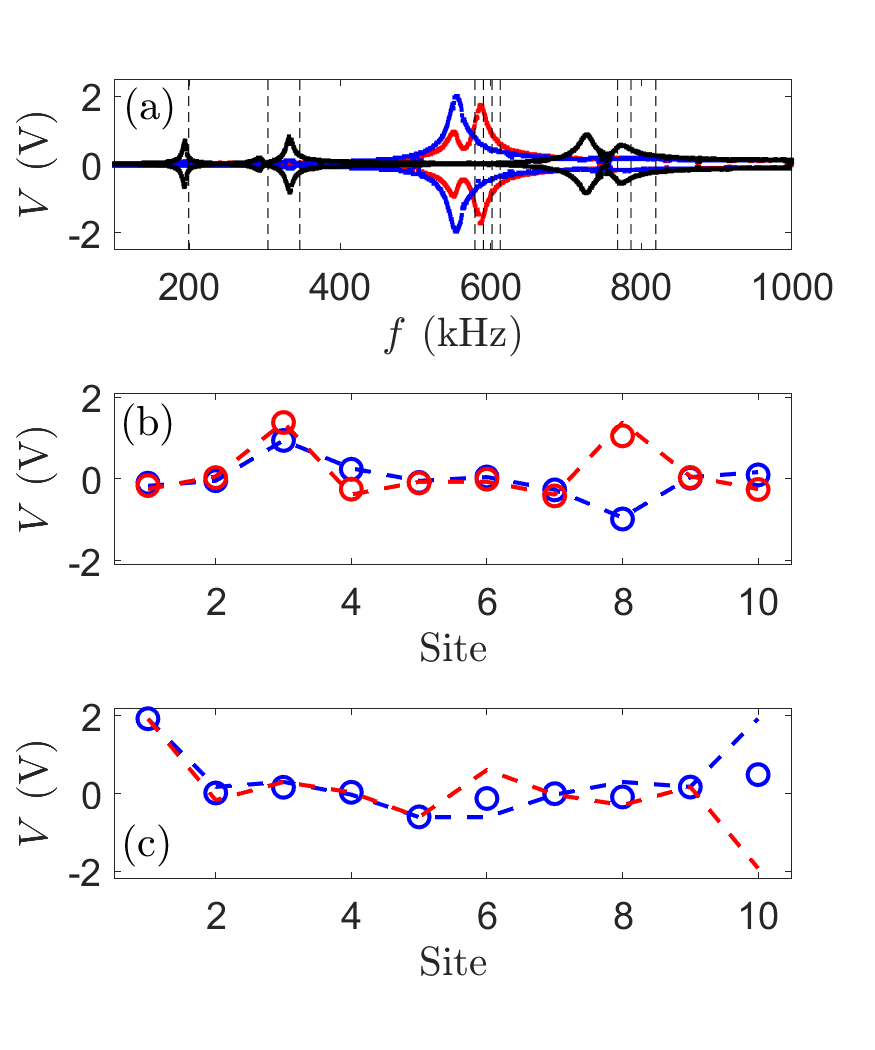}

	\caption{Lattice with long-range coupling between the B-site of the \(n^{\text{th}}\) unit cell and the A-site of the \((n+2)^{\text{th}}\) unit cell. 
		(a) Experimental voltage at the driven node when energy is injected locally at the end of the lattice (blue), at the second node (black), and at the third node (red), using a sinusoidal voltage source of amplitude \(V_d = 5\,\text{V}\) in a configuration where \(\mu = -2\). 
		The dashed vertical lines mark the theoretical resonance frequencies. 
		Panels (b) and (c) display the normalized numerical (dashed lines) and experimental spatial profiles (circles) of the edge modes. 
		A small frequency shift appears due to lattice imperfections and external impedances introduced by the measurement instrumentation. 
		The experimental edge mode localized at the end of the lattice seems to be a superposition of the two theoretical edge modes, which could not be experimentally resolved. 
		(b) \(f_{\text{num}} = 579\,\text{kHz}\) and \(f_{\text{exp}} = 549\,\text{kHz}\) in blue, and  \(f_{\text{num}} = 602\,\text{kHz}\) and \(f_{\text{exp}} = 557\,\text{kHz}\) in red.
		(c) \(f_{\text{num}} = 590\,\text{kHz}\) and \(f_{\text{exp}} = 557\,\text{kHz}\) in blue, and \(f_{\text{num}} = 602\,\text{kHz}\) in red.  
		Circuit parameters: \(L_1 = 0.33\,\text{mH}\), \(L_2 = 1\,\text{mH}\), \(L_3 = 0.1\,\text{mH}\), and \(C = 1\,\text{nF}\).
	}
	\label{scan_profile_lr_a}%
\end{figure}

\begin{figure}[h]
	\centering
	\includegraphics[width=\columnwidth]{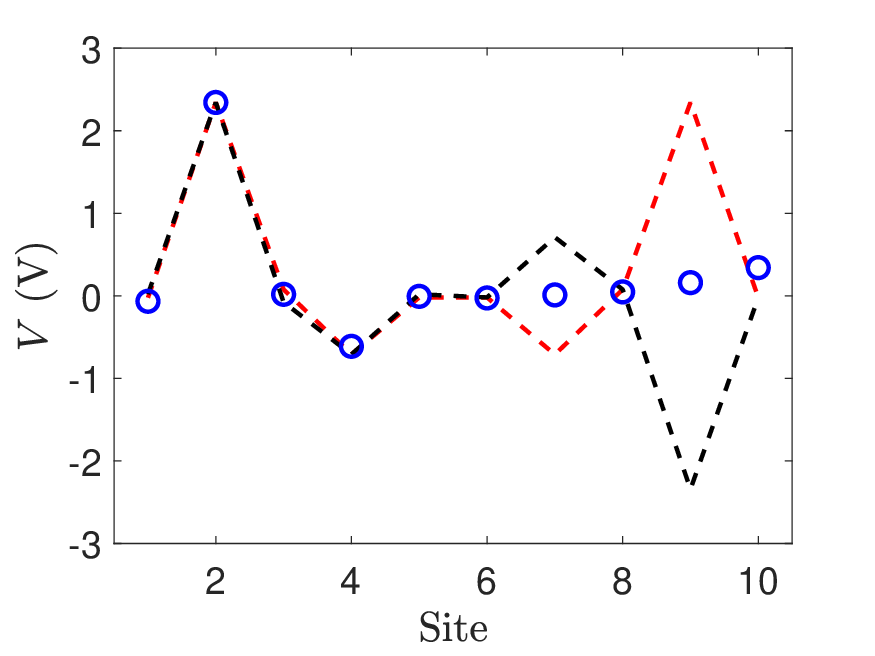}
	\caption{ Normalized numerical (dashed lines) and experimental (circles) spatial profile of the edge modes for a winding number \(\nu = 1\) in a lattice with a long-range coupling corresponding to the interaction between the A-site of the \(n^{\text{th}}\) unit cell and the B-site of the \((n+1)^{\text{th}}\) unit cell. 
	The corresponding frequencies are \(f_{\text{num}} = 596\,\text{kHz}\) and \(f_{\text{exp}} = 543\,\text{kHz}\). 
	Circuit parameters: \(L_1 = 0.33\,\text{mH}\), \(L_2 = 1\,\text{mH}\), \(L_3 = 0.1\,\text{mH}\), and \(C = 1\,\text{nF}\).
	}
	\label{scan_profile_lr_b}%
\end{figure}

In the second configuration, where the A-site of the \(n^{\text{th}}\) unit cell is connected to the B-site of the \((n+1)^{\text{th}}\) unit cell, the system does \emph{not} exhibit a topological phase with winding number \(\pm 2\). Instead, it displays distinct topological phases characterized by winding numbers \(\nu = -1\) and \(\nu = +1\), each associated with clearly distinguishable edge states. These results are summarized in Figs.~\ref{phase_lr}, \ref{scan_profile_lr_a}, \ref{scan_profile_lr_b}. 

The main finding here is that the two implementations of longer-range coupling lead to additional and distinct edge modes. When $\nu=\pm 2$, edge modes can localize on $n=1$ and $n=10$ (the left-most and right-most sites) or on $n=3$ and $n=8$. In the latter case, we see the peaks are close enough in space that the frequency shift between the even and odd solutions is great enough that we can address them separately in the experiment. Thus, even though the driving occurs at site $n=3$, we obtain the even and odd edge modes depending on the precise driving frequency. When driving $n=1$, the frequency shift between those even and odd edge mode solutions is small enough that only the even superposition can be excited.
Finally, for the case of $\nu=1$, we obtain edge modes centered on site $n=2$ and $n=9$, both analytically and experimentally.

Overall, the theoretical predictions, numerical simulations, and experimental measurements exhibit a high degree of consistency. The agreement among theory, numerics, and experiment provides evidence for the robustness of the topological phenomena described.

\section{Bound states at domain walls}
We now turn to the existence of topological edge states at domain walls between different regions of the lattice. To illustrate the situation, we consider the two limiting configurations corresponding to weak coupling ($v \rightarrow 0$, $L_1 \rightarrow \infty$). In one case, the lattice contains an isolated site, while in the other case it forms a trimer. In both situations,  the states exhibit an eigenvalue $\lambda$ of zero and are therefore degenerate in frequency.
\begin{figure}[h]
\centering
\includegraphics[width=\columnwidth]{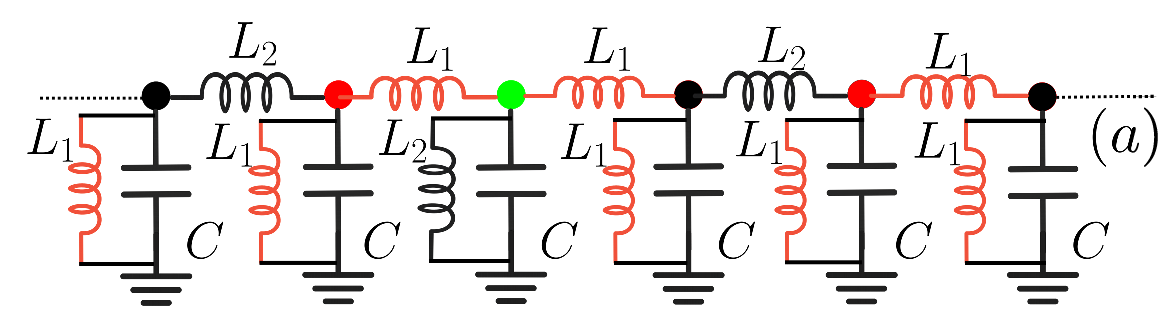}
\includegraphics[width=\columnwidth]{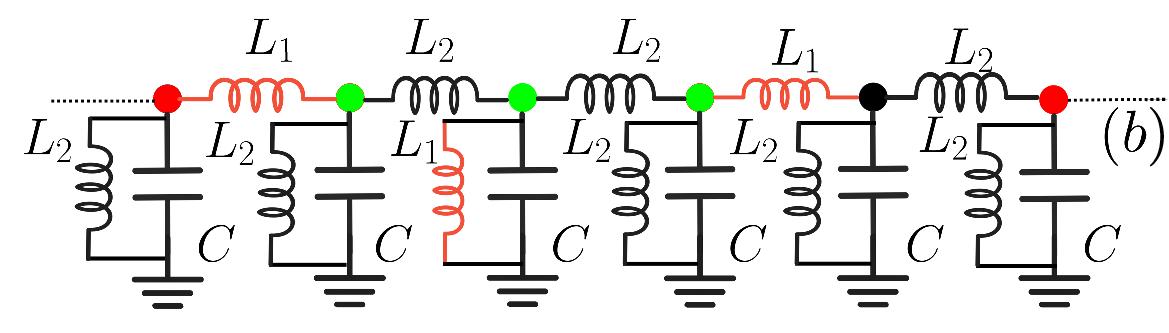}
\caption{Two different domain wall configurations: (a) Monomer case and (b) trimer case. Domain walls are represented by  green circles.}
\label{dw1}%
\end{figure}

To generate these modes in our electrical lattice, we implemented the two configurations shown in Fig. (\ref{dw1}). To preserve the structure of the governing equations, an additional inductor is introduced in parallel with the capacitor, as shown in the lattice circuit diagrams. (In the monomer case, it would, in principle, be sufficient to add only one extra inductor $L_2$ in parallel with the capacitor; however, we adopt this option in both cases to maintain a consistent structure.) Finally, to avoid interference with edge modes at the boundaries, we impose periodic boundary conditions.

\begin{figure}[h]
\centering
\includegraphics[width=\columnwidth]{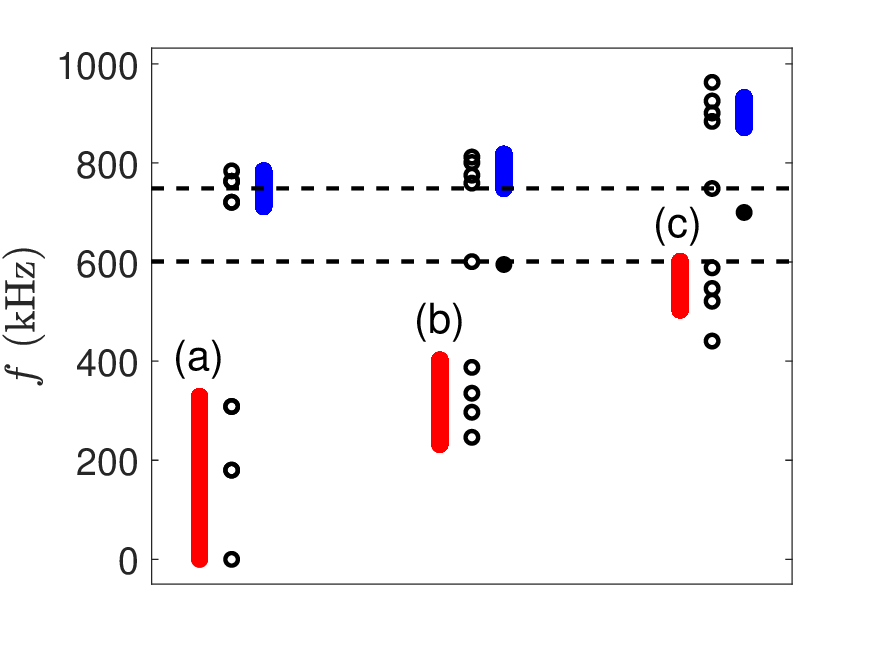}
\caption{Band structure of an electrical lattice with periodic boundary conditions: (a) no domain walls, (b) monomer case, and (c) trimer case. Open black circles represent numerical calculations,  while red (blue) lines correspond to the acoustic (optic) dispersion bands of an infinite lattice without domain walls. Dashed lines indicate the frequencies of edge modes in the infinite lattice. Specifically, these frequencies are $\Omega=\sqrt{v+w}$ for the case without domains, $\Omega=\sqrt{2v+w}$ for the monomer case, and $\Omega=\sqrt{v+2w}$ for the trimer case. Solid black circles represent experimental domain-wall nodes, with $f = 505\,\mathrm{kHz}$ in the monomer case and $f = 700\,\mathrm{kHz}$ in the trimer case. $L_1 = 470,\mu$H, $L_2 = 100,\mu$H, $C = 400$ pF, and 10 nodes (9 in domains walls situation).}


\label{dw_bands}%
\end{figure}

We determine the linear modes corresponding to the different configurations numerically by diagonalizing the corresponding matrix. The resulting band structure is presented in Fig. (\ref{dw_bands}), which clearly displays the presence of a localized mode within the band gap. 
Here, the blue circles and red lines correspond to analytical and numerical results, respectively. 

In the thermodynamic limit ($N \rightarrow \infty$), it is possible to find an analytical expression for the domain wall edge modes by solving the corresponding linear equations. Modes have a frequency $\Omega=\sqrt{2 v+w}$ in the monomer case, and the amplitudes at nodes $n$ and $z_n$ follow the pattern.
\begin{equation}
 z_{n_0 \pm i} =
\begin{cases}
C_1 (-v/w)^{i/2}, & \text{if } i \text{ is a even number} \\
0, & \text{if } i \text{ is an odd number}.
\end{cases}
\label{ap_m}
\end{equation}
The singular node, coupled to its neighbors through inductors $L_1$, is labeled $n_0$. $C_1=\sqrt{(w^2-v^2)/(v^2+w^2)}$ is a normalization constant.

In trimer case, $\Omega=\sqrt{v+2 w}$ and its spatial profile is given by,
\begin{equation}
 z_{n_0 \pm (i+1)} =
\begin{cases}
\pm C (-v/w)^{i/2}, & \text{if } i \text{ is a even number} \\
0, & \text{if } i \text{ is an odd number},
\end{cases}
\label{ap_tri}
\end{equation}
with $C_2=\sqrt{2 w^2/(w^2-v^2)}$ as the normalization constant.

To experimentally detect the domain wall modes, we use a method similar to the one described previously. The lattice is driven sinusoidally through a small capacitor ($C_d = 24$ pF), applied either at the domain wall node in the monomer case or at two neighboring nodes of the domain wall with out-of-phase signals generated by an inverter. By measuring the voltage at the domain wall, we obtain the results shown in Fig. (\ref{dw_bands}). In general, we again find good agreement between numerical and experimental results.

\begin{figure}[h]
\centering
\includegraphics[width=\columnwidth]{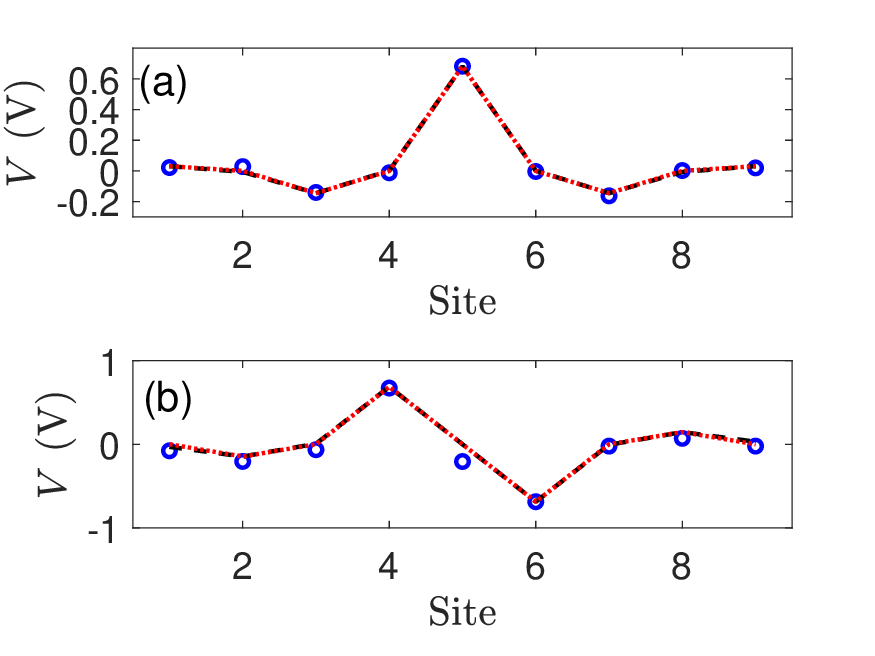}
\caption{(a) Edge mode profiles (a) monomer and (b) trimer. Circles correspond to experimental measurements, dashed black line to normalized numerical calculations and dotted red line to normalized approximate solutions determined by (\ref{ap_m}) and (\ref{ap_tri}). Circuit parameters: $L_1 = 0,470 \,\text{mH}$, $L_2 = 0.1 \,\text{mH}$, and $C = 1 \,\text{nF}$.}
\label{edge_3}%
\end{figure}

The response can be plotted spatially along the lattice, and the results are shown in Fig.~\ref{edge_3}. We see in panel (a) that in the monomer case, the mode energy is centered on the interface site between the two domains. In contrast, for the trimer case, panel (b) reveals the domain-wall state to have odd symmetry about the interface site, representing an energy node. In this figure, blue circles depict experimental data, whereas lines represent numerical and analytical results.      

\section{Conclusions} 

In this paper, we have introduced the main concepts underlying topologically protected states using the simple and pedagogical Su--Schrieffer--Heeger (SSH) model. We have constructed a physical realization of this system using readily accessible electrical components, allowing us to experimentally verify the existence and several key properties of these states. The experimental procedure has been described in detail so that it can be easily reproduced.

These ideas can be extended in several promising directions. One possibility is to design longer lattices and incorporate various forms of long-range coupling, which can give rise to a greater number of topologically localized modes. Another natural extension is the exploration of two- and three-dimensional analogs, where the corresponding edge, surface, or hinge states exhibit richer topological behavior. Another very interesting direction is to incorporate a nonlinear element and investigate how nonlinear effects modify the behavior of the system.

Importantly, the experimental setups required to investigate these systems remain inexpensive, modular, and accessible to a broad range of students and researchers, making them valuable tools for both education and research in topological bound states.

\section{Acknowledgments}
Financial support from the Ministerio de Ciencia, Innovación y Universidades (MICIU, Spain) for mobility stays in foreign higher education and research institutions is gratefully acknowledged (F.P.). F.P. also wishes to express his sincere gratitude to Dickinson College for its kind hospitality.

\section{Declaration of generative AI and AI-assisted technologies in the manuscript preparation process}
During the preparation of this work the authors used the language-editing assistance of OpenAI’s ChatGPT in order to improve grammar, wording, and syntax in portions of the manuscript. The tool was not used for generating scientific content, analysis, or results. After using this tool/service, the authors reviewed and edited the content as needed and takes full responsibility for the content of the published article.

\end{document}